# THERMAL HISTORY CONSTRAINTS ON THE ISOCURVATURE BARYON MODEL [†]

Wayne Hu[1] and Naoshi Sugiyama[1,2]

[1]*Departments of Astronomy and Physics*
*University of California, Berkeley, California 94720*

[2] *Department of Physics, Faculty of Science*
*The University of Tokyo, Tokyo, 113, Japan*

We present a thorough and detailed investigation of baryon isocurvature models including realistic thermal histories, *i.e.*, late or partial reionization of the universe and compact object formation after standard recombination. Constraints on these models are imposed from spectral distortion in the cosmic microwave background (CMB), number fluctuations in galaxy counts $\sigma_8$, and recent CMB anisotropy observations. Since the status of degree scale anisotropies is as yet unclear, the lack of spectral distortions is the most serious constraint on the spectral index of initial fluctuations: $-1.2 \lesssim n \lesssim -0.5$ in partially ionized models and $-1.2 \lesssim n \lesssim -0.9$ in compact object dominated universes in order to avoid significant bias in galaxy formation. Full ionization of the universe before $z = 300$ is also forbidden by this constraint. Intermediate scale CMB fluctuations which are significantly larger than standard CDM is a *robust* prediction of these models. Many models are consequently in conflict with the recent degree scale detections by Python, ARGO, and MSAM, *e.g.*, compact object dominated models with $\Omega_0 \lesssim 0.2$. Moreover, *all* models will be ruled out if the low CMB fluctuations on degree scales as detected by the SP91 13 point scan are confirmed. On the other hand, most models fare well compared with the high Tenerife and MAX detections.

*Subject Headings:* Cosmology: Cosmic Microwave Background, Cosmology, Theory

hu@pac2.berkeley.edu, sugiyama@bkyast.berkeley.edu

*My life flows between confines*
*But knowledge has no confines.*
*If we use the constrained*
*To follow after the unconstrained,*
*There is a danger that the flow will cease;*
*And when it ceases,*
*To exercise knowledge is purest danger.*
       *Chuang-tzu*

## 1. INTRODUCTION

The isocurvature baryon model is one of only a handful of explanations for structure formation in the universe (Suginohara & Suto 1992; Cen, Ostriker & Peebles 1993). This model was proposed by Peebles (1987) to satisfy dynamical measurements of a low $\Omega_0 \approx 0.1 - 0.2$ (*e.g.* Peebles 1986, Dekel *et al.* 1993) while simultaneously avoiding the need for hypothetical non-baryonic dark matter or a non-zero cosmological constant. Recent observations of microlensing from compact halo objects may also support the idea of baryonic dark matter (Alcock *et al.* 1993). With the COBE DMR detection of large angle anisotropies in the Cosmic Microwave Background (CMB) and the COBE FIRAS upper limit on Compton-$y$ spectral distortions, however, the isocurvature baryon model with the simplest and most plausible thermal histories is ruled out. Normalized to the COBE DMR detection at $10°$ of $(\Delta T/T)_{rms} = 1.12 \pm 0.10 \times 10^{-5}$ (Bennett *et al.* 1994), the isocurvature baryon model with a standard recombination history vastly overproduces temperature fluctuations just under the degree scale. In this model, however, structure can form immediately after recombination, and an early reionization is plausible (Peebles 1987). Nevertheless, although rescattering damps the small scale temperature fluctuations to allowable levels, it also gives rise to unacceptably high spectral distortions in a fully ionized model (Gnedin & Ostriker 1992, Tegmark & Silk 1994). Therefore, only models with a non-standard thermal history, *e.g.*, a partially or recently ionized universe, can survive these opposing constraints.

Moreover, it is by no means obvious that one can construct a model consistent with CMB observations at the degree scale even allowing for an *ad hoc* ionization history. Recently, Chiba, Sugiyama, & Suto (1994) have shown that the fully ionized isocurvature baryon model is inconsistent with the SP91 13 point scan detection at $\sim 2°$ (Schuster *et al.* 1993). Indeed we show here that if low fluctuations at this scale are confirmed, *all* isocurvature baryon models are excluded regardless of ionization history or even the possible formation of compact objects such as black holes. However, several recent CMB experiments, *e.g.* Python (PYTH, Dragovan *et al.* 1993), MAX GUM (Gundersen *et al.* 1993), MSAM (Cheng *et al.* 1993), have all reported higher detections than SP91 on similar angular scales. Moreover, the second year of COBE DMR data and the high detection by Tenerife (TENE) at the $5°$ scale (Hancock *et al.* 1993) indicate a steeper than Harison-Zel'dovich spectral slope at large scales. The standard isocurvature baryon model, which has low $\Omega_0$, can naturally account for this effect (Sugiyama & Silk 1994). Furthermore, since large scale fluctuations are nearly independent of ionization history, all models considered here can also account for this feature.

Since the data at scales smaller than the COBE DMR detection are marginally consistent at best, it is no surprise that no isocurvature baryon model can satisfy all the constraints from the data. Until this situation is resolved, we cannot judge the viability of this model on the basis of intermediate scale anisotropies. We do however present the full CMB predictions of these models which can be employed once the measurements improve. In light of this current uncertainty, we also place constraints on the thermal history of this model from first and second order temperature fluctuations at arcminute scales, spectral distortions, and galaxy clustering. We provide the first consistent treatment of the matter power spectrum in compact object dominated universes and find significant differences from other works (*e.g.* Cen, Ostriker, & Peebles 1993), which may have undesirable consequences for large scale structure formation. The favored values for the slope of the power spectrum $n \gtrsim -0.5$ also overproduce galaxy clustering in the absence of antibias, if consistency with spectral distortions is also required.

The plan of this paper is as follows. In §2, we explore the dependence of matter and temperature fluctuations, as well as spectral distortions, on thermal history including compact object formation. This allows us to place constraints and make predictions for the various models in §3. In §4, we discuss the implications of these results and evaluate the present status of the isocurvature baryon scenario with respect to CMB anisotropies.

## 2. THERMAL HISTORY EFFECTS

First, we briefly summarize the general properties of the isocurvature baryon model. This model consists of only baryons, photons and three massless neutrinos in a low density open universe. Initially, there are no fluctuations in the total density, but rather perturbations in the entropy per baryon. We thus assume initial conditions such that this entropy fluctuation follows a pure power law in $\tilde{k}$, $S^2(\tilde{k}) \propto \tilde{k}^n$ where $\tilde{k}^2 = k^2 + K$ and $K = -H_0^2(1 - \Omega_0)$ (Wilson 1983). In such models, the absence of Silk damping on small scales allows for the collapse of objects immediately following standard recombination at $z \approx 1000$. This first generation of objects could reionize the universe to a significant degree (Peebles 1987). It is also possible that a large fraction of the mass ends up in these compact objects and thereupon behaves effectively as cold dark matter (Gnedin & Ostriker 1992). Gnedin & Ostriker also propose that compact object formation may alter nucleosynthesis and resolve the most serious difficulty of this model: $\Omega_0 = \Omega_B \gg 0.015h^{-2}$, the value required by standard nucleosynthesis (Smith *et al.* 1993). A second phase of structure formation, perhaps associated with galaxy formation, could further ionize the universe. Therefore the following parameters adequately describe the range of possibilities for the model: $z_c$, the epoch of compact object formation; $x_c$, the ionization fraction after $z_c$; $\Omega_{IGM}$, the amount of matter left in the intergalactic medium; $z_i$ the secondary ionization redshift; and $x_i$, the ionization fraction after secondary ionization. We will now describe the effects that these parameters have on the evolution of perturbations in the matter and radiation. For a more detailed treatment of these effects, see Hu & Sugiyama (1994).

### 2.1 The Matter Power Spectrum

Let us first consider the simplest case where the effect of compact objects is negligible and reduce our parameters to $x_i$ and $z_i$. Gravitational instability and microphysical processes will transform the initial perturbation into a power spectrum of the form $P(k) = (T(k)S(k))^2$. We solve the first order Boltzmann equations for the coupled photon-baryon system as well as the neutrinos, to the present, in order to obtain the transfer function $T(k)$ (Sugiyama & Gouda 1992, Hu & Sugiyama 1994).

Even with isocurvature initial conditions, a curvature perturbation will be stimulated in order to keep the entropy constant (Kodama & Sasaki 1986). These perturbations will grow as the adiabatic mode and become important as the perturbation enters the horizon in the matter dominated epoch. However, when the adiabatic component enters the Jeans length, pressure gradients will force it to oscillate and damp away. Since entropy is conserved in the tight coupling limit, on small scales the baryons are left with the initial entropy perturbation. On large scales, the initial entropy perturbation is shifted from the matter to the radiation as the universe becomes matter dominated, in order to avoid a large curvature perturbation on superhorizon scales. Thus the matter is left with a characteristic $k^2(1 - 3K/k^2)$ tail to the transfer function due to the remaining fluctuations from the adiabatic growth.

The peak of the transfer function therefore corresponds to the maximal Jeans scale (see Fig. 1). For models in which standard recombination is followed by a significant transparent period, the peak will be close to the Jeans scale at recombination, $k_J^{(SR)} \approx 0.13[1 + 0.24(\Omega_0 h^2)^{-1} + 0.007(\Omega_0 h^2)^{-2}]^{1/2} \Omega_0 h^2 \mathrm{Mpc}^{-1}$, and the oscillations at small scales will not have had a chance to damp away. For a universe that never recombines (or $z_i \approx 1000$), the maximum Jeans length is larger, $k_J^{(NR)} \approx 0.13 \Omega_0 h^2 \mathrm{Mpc}^{-1}$, since standard recombination is close to matter-radiation equality in low $\Omega_0$ models. Between standard recombination and $z_i$, fluctuations grow in linear theory, and thus will be larger for models with low $z_i$. On the other hand, from $z_i$ to the end of the drag epoch $z_{drag} \approx 160(\Omega_0 h^2)^{1/5} x_i^{-2/5}$, Compton drag will again prevent fluctuations from growing inside the Jeans length. Since Compton drag is less effective for low $x_i$, these models have larger fluctuations at the present and a less prominent peak. If claims that the observational power spectrum is quite smooth are confirmed (Peacock & Dodds 1994), then these models may be the only viable ones.

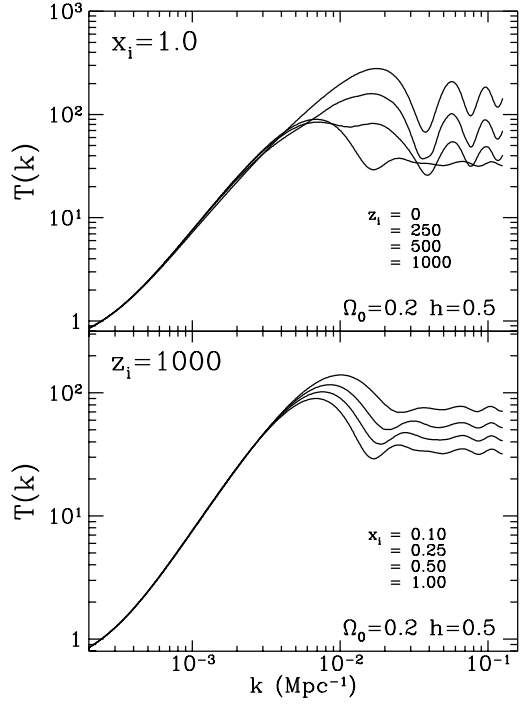

FIG. 1. Matter transfer function in partially ionized models. Scenarios with low $z_i$ have their peak on smaller scales due to the smaller Jeans length at standard recombination. Those with high $x_i$ suffer suppression in growth due to high Compton drag. Compact object dominated scenarios will resemble the standard recombination model of $z_i = 0$.

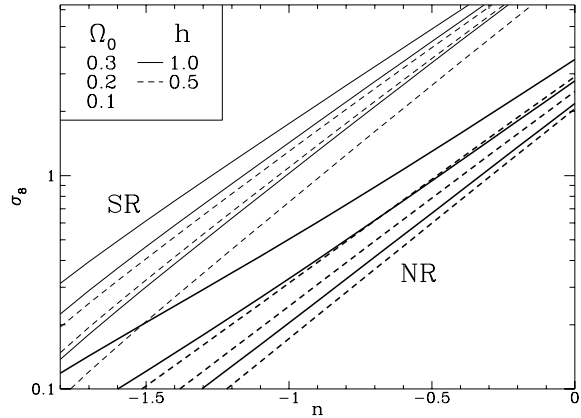

FIG. 2. Present day amplitude of fluctuations on the $8h^{-1}$Mpc scale $\sigma_8(n)$ for a normalization to the COBE DMR detection at $10°$. Fluctuations in the no recombination (NR, thick) scenario are suppressed with respect to the standard recombination counterparts (SR, thin) due to Compton drag. The value of $\Omega_0$

decreases from top to bottom for $h = 1.0$ (solid line) and $h = 0.5$ (dashed line). All partially ionized scenarios have fluctuations in between these limits. Models which are compact object dominated have nearly the (SR) values. Analytic fits to these numerical results are found in Tab. 1.

The present amplitude of matter fluctuations for all partially ionized scenarios will therefore lie in between their standard recombination (SR) and no recombination (NR) values as displayed in Fig. 2. All models are normalized to the COBE DMR detection at $10°$. As is readily apparent from this figure, the value of $\sigma_8$, the present day mass overdensity on the $8h^{-1}$ Mpc scale, fits a function of the form $\sigma_8(n) = ae^{bn}$ where the constants $a$ and $b$ are given for standard and no recombination in Tab. 1. The value of $\sigma_8$ for a specific $n$ must fall within the range given by Fig. 2. Moreover, in an unbiased scenario of galaxy formation, $\sigma_8 \approx 1$. Numerical simulations (Cen, Ostriker, & Peebles 1993) indeed seem to indicate that bias at $8h^{-1}$Mpc is low, $b \approx 1.1$. Thus, for $0.1 < \Omega_0 < 0.2$ and $0.5 < h < 1.0$, this implies that the spectral index must lie within the range $-1.2 \lesssim n \lesssim -0.3$ for *any* ionization history. The constraint for a *specific* ionization history is of course much more stringent. Growth during the period where the universe is transparent can be incorporated in a function which is nearly independent of the index $n$,

$$\sigma_8(z_i, x_i) \approx \sigma_8^{(NR)} F(z_i, x_i). \tag{1}$$

We will give the numerical values of this function for various $\Omega_0$ and $h$ in §3. A small $n$ dependence arises at the SR extreme due to the change in the COBE DMR normalization with ionization history.

| $\Omega_0$ | $h$ | $a^{(NR)}$ | $b^{(NR)}$ | $a^{(SR)}$ | $b^{(SR)}$ |
|---|---|---|---|---|---|
| 0.1 | 0.5 | 2.09 | 2.50 | 9.50 | 2.55 |
| 0.1 | 1.0 | 2.16 | 2.35 | 11.3 | 2.42 |
| 0.2 | 0.5 | 2.53 | 2.34 | 11.7 | 2.39 |
| 0.2 | 1.0 | 2.64 | 2.06 | 12.9 | 2.22 |
| 0.3 | 0.5 | 2.94 | 2.23 | 12.5 | 2.28 |
| 0.3 | 1.0 | 3.27 | 1.85 | 13.7 | 2.07 |

TAB. 1. Fitting parameters for $\sigma_8(n) = ae^{bn}$ for standard recombination (SR) and no recombination (NR). These values can be used in conjunction with Fig. 7 to obtain $\sigma_8$ for partially ionized models. In compact object dominated models, $\sigma_8$ is nearly the SR value.

### 2.2 Radiation Anisotropies

Now let us consider the radiation power spectrum. We will separate the contributions to the final anisotropy into *primary* fluctuations, generated at standard recombination, and *secondary* fluctuations, generated afterwards. Doppler shifts from scattering off electrons in small scale bulk motion will create exceedingly large temperature inhomogeneities on the scattering surface at recombination, due to the large amount of small scale power in these models. As the radiation then free streams, the inhomogeneities become anisotropies under the horizon scale and give rise to primary temperature anisotropies on scales of several arcminute to degrees today. When the universe reionizes at $z_i$, Thomson scattering off electrons isotropizes the photon distribution and reduces these small scale primary fluctuations to

$$\left(\frac{\Delta T}{T}\right)_{prim} = \left(\frac{\Delta T}{T}\right)_{SR} e^{-\tau}. \tag{2}$$

Here
$$\tau = \int dt x_i n_e \sigma_T c$$
$$\approx 0.046 x_i h \Omega_{IGM} \Omega_0^{-2} \left[2 - 3\Omega_0 + (\Omega_0 z_i + 3\Omega_0 - 2)\sqrt{1 + \Omega_0 z_i}\right]$$

is the optical depth due to Thomson scattering, where $n_e$ is the electron number density and $\sigma_T$ is the Thomson cross section. For scales above the horizon, the fluctuations are still inhomogeneities, and being isotropic, do not damp due to rescattering. Requiring that the anisotropies satisfy the upper limit on arcminute scales from the OVRO experiment $(\Delta T/T)_{OVRO} < 2.1 \times 10^{-5}$ (Readhead et al. 1989), we set a lower bound on the optical depth $\tau_{min}$ for any given model. In Fig. 3, we plot this minimal optical depth. Notice that it in all cases $\tau_{min}$ is of $\mathcal{O}(1)$ and is only weakly dependent on $\Omega_0$.

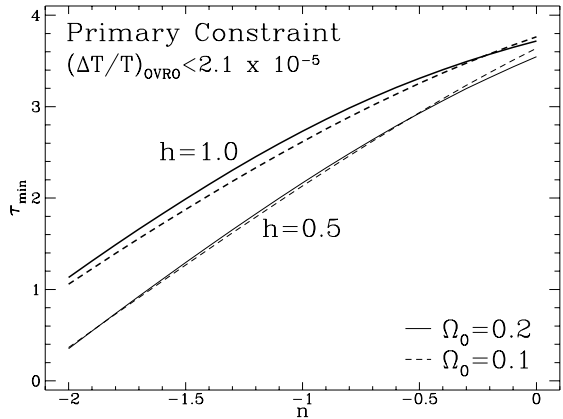

FIG. 3. Primary constraint from OVRO as a function of spectral index $n$. Thick lines represent $h = 1.0$ whereas thin lines denote $h = 0.5$ for two different values of $\Omega_0$, $\Omega_0 = 0.1$ (dashed) and $\Omega_0 = 0.2$ (solid). There is a minimal optical depth necessary to damp primary anisotropies from standard recombination to levels under the OVRO upper limit of $(\Delta T/T)_{OVRO} < 2.1 \times 10^{-5}$. The minimal optical depth is only a weak function of $\Omega_0$.

Up until now we have neglected secondary anisotropies which are generated after standard recombination. Many effects contribute to the final anisotropy: the Doppler effect on the new last scattering surface, the Vishniac (second order Doppler) effect (Ostriker & Vishniac 1986, Vishniac 1987), the adiabatic growth of intrinsic photon fluctuations, the ordinary Sachs-Wolfe effect on the new last scattering surface, the entropy effect, and the integrated Sachs-Wolfe effect (both the curvature and entropy components) (Sachs & Wolfe 1967). The last four effects can be treated in a simple and unified way by examining the Boltzmann equation for the photons (Hu & Sugiyama 1994). We will refer to the combination of these four effects as the *total* Sachs-Wolfe effect.

For the Doppler type effects, the amplitude and shape of the spectrum depend sensitively on when last scattering occured, which is itself determined by $x_i$. On scales smaller than the horizon at last scattering, the Doppler effect is damped due to the cancellation of redshifts and blueshifts from photons that scattered from the front and back of a perturbation. Since the horizon grows with time, this damping effect, due to the finite thickness of the last scattering surface, is maximized by having the most recent last scattering, i.e., the maximal $x_i$. In Fig. 4, we plot the effects of varying $x_i$ on the temperature anisotropies. Note that the $C_\ell$'s are related to the ensemble average of the temperature fluctuations as

$$\left(\frac{\Delta T}{T}\right)^2 = \sum_\ell \frac{2\ell + 1}{4\pi} C_\ell W_\ell \qquad (3)$$

where $W_\ell$ is the experimental window function taken from White, Scott, & Silk (1994). In general, increasing $x_i$ moves the peak of the temperature distortions to larger scales and decreases the amplitude. There is a small effect that can increase the amplitude however. In a highly ionized universe, the drag epoch is somewhat earlier than the last scattering epoch, defined as the time when optical depth reaches unity:

$z_{drag} > z_{ls} \approx 30(\Omega_0/0.1)^{1/3}(0.05/x_i\Omega_{IGM}h)^{2/3}$. The growth of velocities between the two epochs can lead to *slightly* larger temperature fluctuations due to the Doppler effect. As one can see from Fig. 4, this is a very small effect.

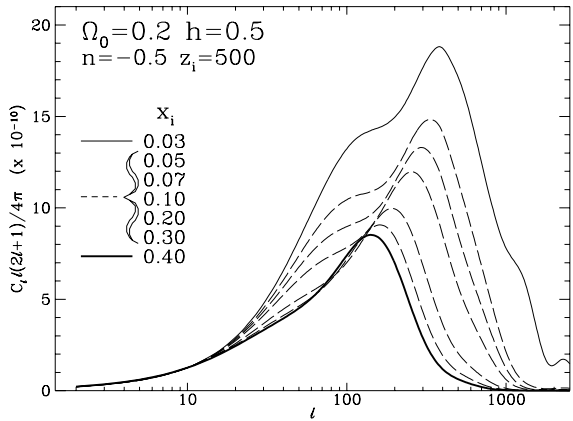

FIG. 4. Dependence of temperature anisotropies on the ionization fraction $x_i$ after the reionization epoch $z_i = 500$. Fluctuations are damped more severely and on larger scales in high $x_i$ models due to the increasing thickness of the last scattering surface. The peak of fluctuations correspondingly moves to larger scales. Aside from a small effect due to growth between the drag epoch and last scattering, raising $x_i$ will always minimize fluctuations.

Since the first order effect is strongly damped at sub-horizon scales, second order effects can play a role (Ostriker & Vishniac 1986, Vishniac 1987). Mode coupling between the density perturbations and the bulk velocity, called the Vishniac effect, can give rise to a significant contribution on arcminute scales. Moreover, it has been shown that this is the dominant second order effect (Hu, Scott, & Silk 1994). However since it is proportional to the square of the density fluctuations, it is only important if perturbations have grown to a significant level by last scattering. This implies that the second order effect will decrease rapidly as one lowers $x_i$ to make the last scattering earlier. We shall quantify this constraint in §3 by using the techniques developed by Efstathiou (1988) and extended by Hu, Scott, & Silk (1994) and Chiba, Sugiyama, & Suto (1994) to open universes.

The integrated part of the total Sachs-Wolfe effect (Sachs & Wolfe 1967), which occurs since the photons are traveling through a time dependent potential, can also play a role at small scales for fully ionized models (Hu & Sugiyama 1993). The relative importance of the small scale effect decreases in partially ionized universes due to an increase in the Doppler term. On the other hand, at large scales the total Sachs-Wolfe effect is the dominant term in all cases. Since such fluctuations are always above the Jeans length, this effect is independent of the ionization history. Thus, the low multipole moments of the temperature anisotropy are nearly identical for all models. This then implies that the COBE DMR normalization is essentially fixed. It is also interesting to note that the slope of the spectrum at large scales is only weakly dependent on $\Omega_0$ and $n$ (Sugiyama & Silk 1994). Compared with the spectral index of flat adiabatic models $n_{ad}$, low $\Omega_0$ isocurvature models predict $n_{ad} \approx 2$ in contrast with the inflationary prediction of $n_{ad} \approx 1$. Recent indications of a steep adiabatic slope in the COBE DMR maps of $n_{ad} = 1.59^{+0.49}_{-0.55}$ (Bennett *et al.* 1994) may argue in favor of this feature.

On intermediate scales, *i.e.*, scales near the horizon at last scattering, the adiabatic growth of the coupled photon-baryon system plays a role. As mentioned above, a curvature perturbation which grows as the adiabatic mode will dominate as the fluctuation comes within the horizon. Therefore, the monopole component of the photon distribution, *i.e.*, the energy density fluctuation in the photons, will grow with the baryons *if* the universe is ionized. After last scattering, this perturbation is transferred to anisotropies at $\ell \sim 100$. Hence at intermediate scales, the photon fluctuations will be the largest for early ionization. The opposite is true for the small scales at which the Doppler effect dominates. For early ionization, the baryon velocity cannot grow due to Compton drag. Therefore at smaller scales, fluctuations will be lower for early ionization. In Fig. 5 we display the dependence of the anisotropies on $z_i$. We have fixed $x_i = 0.1$ to isolate this effect. In all these models, the last scattering surface consequently has the same redshift and thickness. As one can see, intermediate scale fluctuations are minimized by having low $z_i$.

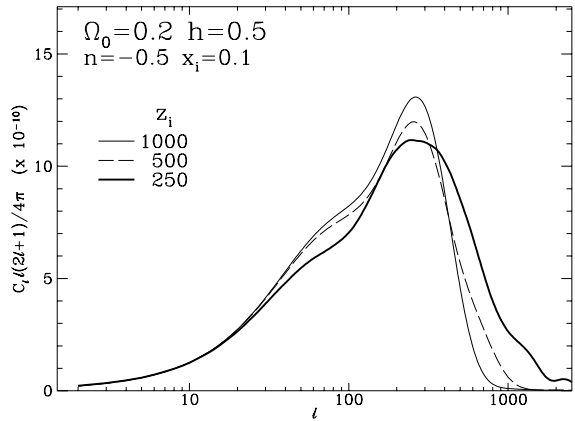

FIG. 5. Dependence of temperature anisotropies on the reionization epoch $z_i$ with fixed last scattering surface ($x_i = 0.1$ $z_{ls} \approx 110$). In models with high $z_i$, the photon fluctuations have had longer to grow adiabatically with the matter on large scales ($\ell \lesssim 100$). On small scales, velocities are prevented from growing by Compton drag, and thus the Doppler effect is smaller for high $z_i$ models.

### 2.3 Spectral Distortions

Ionization also produces spectral distortions by the SZ effect (Zel'dovich & Sunyaev 1969) in all reasonable scenarios, since the intergalactic medium must have been heated by some process such as collisional ionization. A conservative estimate of the necessary temperature is $T_e \gtrsim 5000K$ (*e.g.* Cen, Ostriker, & Peebles 1993 find $T_e \approx 10000 - 20000K$). As CMB photons scatter off this hot medium, a spectral distortion of the Compton-$y$ type will be established as low frequency photons are shifted to higher frequencies. The Compton-$y$ parameter is given by

$$y = \int \frac{k(T_e - T)}{m_e c^2} \dot{\tau} dt \qquad (4)$$

where the overdot is a time derivative, $T = T_0(1 + z)$, and $T_0 = 2.726 \pm 0.005$ is the present temperature of the CMB (Mather *et al.* 1994). For constant $T_e$, the integral is approximately,

$$y \approx 8.4 \times 10^{-7} \tau \left(\frac{T_e}{5000K}\right) \left(1 - \frac{3}{5}\frac{T_0}{T_e} z_i\right) \qquad (5)$$
$$< 2.5 \times 10^{-5},$$

where the constraint comes from the COBE FIRAS experiment (Mather *et al.* 1994). For $T_e = 5000K$, the FIRAS limit becomes $\tau \lesssim 30/(1 - 0.003 z_i)$. Combined with the minimal optical depth needed to suppress primary fluctuations (see equation (2)), the ionization history is severely constrained.

### 2.4 Compact Objects

Now let us consider the effect of having a significant fraction of the universe in compact objects that behave essentially as cold dark matter. The clustering of compact objects is independent of the ionization history. Thus, if the compact objects dominate dynamically over baryons in the intergalactic medium, *i.e.* $\Omega_C \gg \Omega_{IGM}$, the baryons would have fallen into the compact object wells after the drag epoch. If the universe is transparent between recombination and compact object formation, the transfer function will be nearly identical to the standard recombination one. Note that the power spectrum in these models will have prominent oscillations on small scales. The peak will also be at smaller scales than for the partially ionized models. Prior treatments have on the contrary assumed that the power spectrum is identical to the fully or partially ionized spectrum (Cen, Ostriker, & Peebles 1993). In these models, $\sigma_8$ also takes on the value of the standard recombination result displayed in Fig. 2.

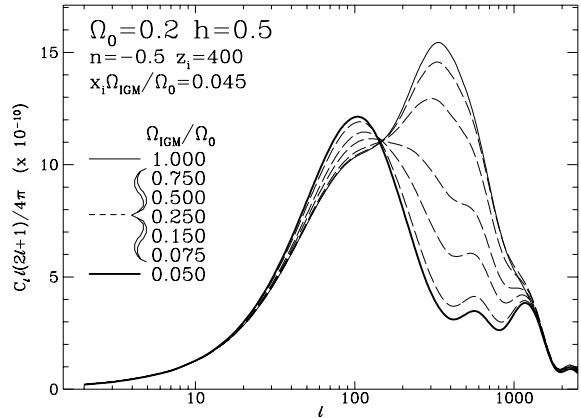

FIG. 6. Dependence of temperature anisotropies on compact objects $\Omega_C = \Omega_0 - \Omega_{IGM}$. In all models, compact objects form at $z_c = 800$, and the universe is transparent until $z_i = 400$. For a fixed optical depth $x_i \Omega_{IGM}/\Omega_0 = 0.045$, raising $\Omega_C$ makes drag effects on the baryons larger. On large scales the photon "velocity" (*i.e.* dipole) is larger than the baryon velocity. The converse is true on small scales. Thus the baryon velocity will be kicked higher at large scales and dragged lower on small scales. The anisotropies suffer a similar effect due to feedback through the Doppler effect.

The behavior of the radiation fluctuations is more complicated however. First of all, lowering the density of matter in the intergalactic medium makes last scattering earlier thereby creating larger fluctuations at small scales (see Fig. 4). To isolate the additional effects of compact objects, we will fix the optical depth $\tau(z)$ in these models by requiring that $x_i \Omega_{IGM}$ and $z_i$ be the same for all models (see Fig. 6). This fixes the location of the last scattering surface and the suppression of the primary anisotropies. However, the Compton drag on any given baryon is independent of $\Omega_{IGM}$ and is larger for models with high $x_i$. If the baryon velocity is greater than the photon "velocity" (dipole), then Compton drag will decrease the velocity. On the other hand, if the photon velocity is larger, the drag will turn into a kick and increase it.

Before recombination, the baryon and photon velocities are comparable due to tight coupling. During the free streaming epoch between recombination and reionization, on scales smaller than the horizon at recombination, the matter perturbations have already joined into the adiabatic growing mode and will just grow in linear theory. The photon dipole however damps due to free streaming. Thus, at small scales the photon velocity at reionization is smaller than the baryon velocity. On larger scales, the baryon velocity decays slightly after recombination as it attempts to join the growing mode of adiabatic perturbations. Thus, the photon velocity at large scales is larger than the baryon velocity. Since last scattering happens relatively near the drag epoch in partially ionized models, there is a corresponding feedback to the temperature anisotropies due to the Doppler effect. Models with lower $\Omega_{IGM}$ (higher $\Omega_C$) will therefore experience larger fluctuations at large scales $\ell \lesssim 100$ due to the kick imparted by the photons and smaller fluctuations at small scales due to the drag associated with the photons. Notice that in Fig. 6, there is also a contribution on very small scales $\ell \gtrsim 1000$ due to the primary anisotropy. All models show the same amount of primary fluctuations since we have fixed the optical depth (see equation (2)). To minimize fluctuations at $\ell \lesssim 100$, drag must be minimized i.e., for fixed optical depth we want *no compact objects*.

In summary, raising $x_i$ will uniformly decrease anisotropies, whereas lowering $z_i$ will decrease anisotropies at the intermediate ($\ell \sim 100$) scale and increase them at the small scale. Furthermore, $x_i$ and $z_i$ must be low enough to satisfy constraints on spectral distortions (see eqn. (4)) but high enough to damp the primary fluctuations (see eqn. (2)). Since these dual constraints limit the range of possible optical depths, the addition of compact objects will have the added effect of raising intermediate scale fluctuations and lowering small scale fluctuations. Furthermore, since minimal fluctuations are generated by the latest allowable last scattering, all compact object models have intermediate scale temperature anisotropies larger than this minimal model. The relative complexity of this picture is just a function of the number of free parameters in this theory and unfortunately cannot be avoided.

## 3. CONSTRAINTS AND PREDICTIONS

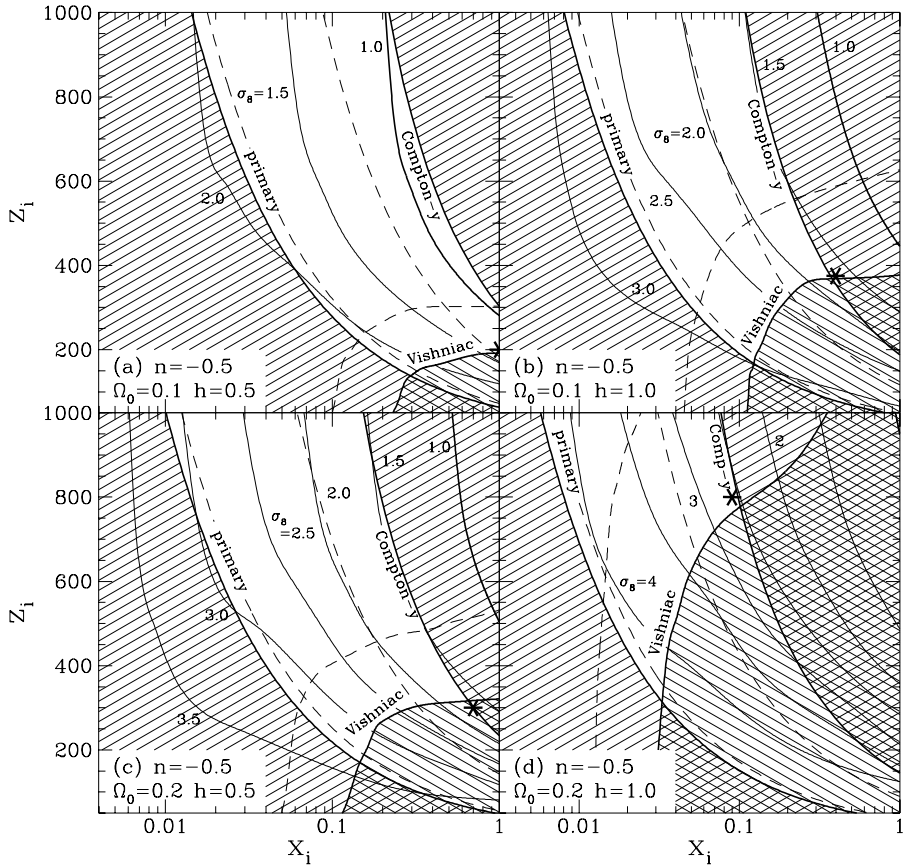

FIG. 7. Constraint plane for the ionization history with $n = -0.5$. The "primary" constraint arises from requiring the standard recombination anisotropies be damped sufficiently to escape the OVRO limit of $(\Delta T/T)_{OVRO} < 2.1 \times 10^{-5}$ on the scale of several arcminutes (see equation (2) and Fig. 3). The Compton-$y$ constraint assumes a conservative minimal temperature for ionization of $T = 5000$K. The Vishniac constraint from the ATCA experiment $(\Delta T/T)_{ATCA} < 0.9 \times 10^{-5}$ on arcminute scales constrains models with recent last scattering and high normalization. Dashed lines represent the constraints if the upper limits of the corresponding experiment are improved by a factor of 2. Approximate contours of constant $\sigma_8$ are given (thin solid lines, with $\sigma_8 = 1$ bolded), which also define the ionization function $F(z_i, x_i)$ (see equation (1) and text). The value of $\sigma_8$ for any $n$ can be generated with the aid of Fig. 2 or Tab. 1. Minimal fluctuations at $\ell \lesssim 100$ are given by the model with largest allowable $x_i$ and lowest allowable $z_i$. The models of Fig. 9, which are essentially these minimal models, are marked out with an asterisk on this figure. Corresponding temperature anisotropies are given in Fig. 9 and Tab. 2.

We will now quantify the considerations of the previous section in order to constrain the ionization history, spectral index, and the formation of compact objects and make model predictions for various degree scale anisotropy experiments. As discussed above, the constraint from primary fluctuations and the Compton-$y$ parameter work in opposite senses: the former requires high optical depth, the latter low. The Vishniac effect constrains models with a high amplitude of present fluctuation (large $\sigma_8$) and late last scattering. The ATCA measurement on arcminute scales of $(\Delta T/T)_{ATCA} < 0.9 \times 10^{-5}$ (Subrahmanyan et al. 1993) severely constrains high $\Omega_0$ and $h$ models through this second order effect (Hu, Scott, & Silk 1994).

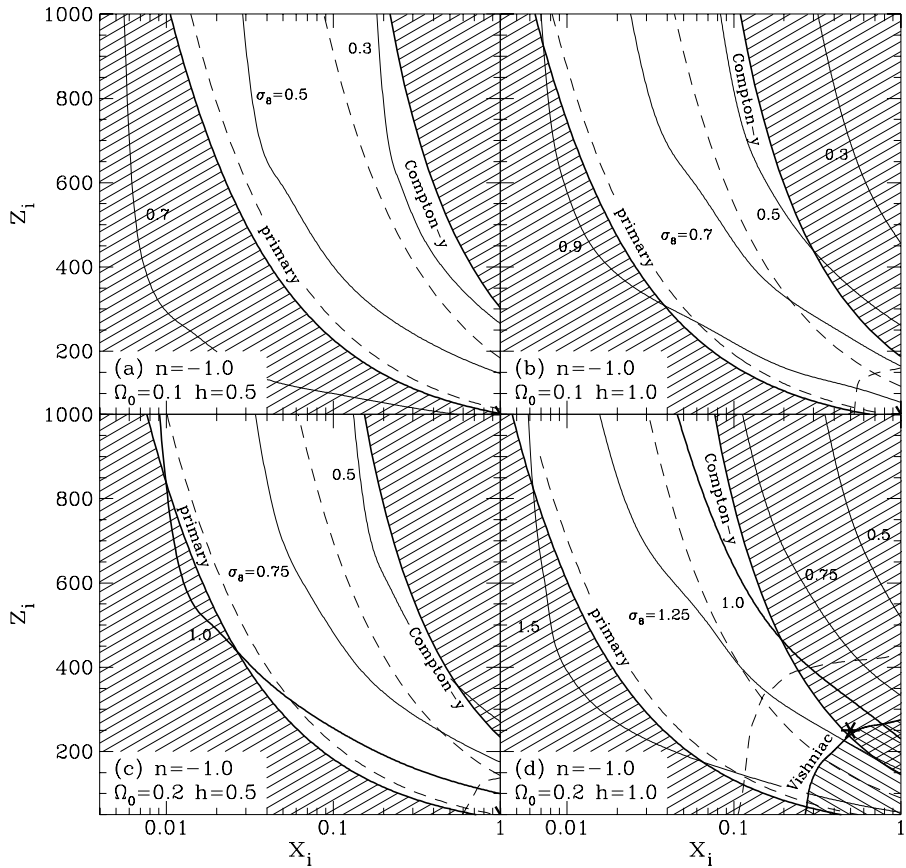

FIG. 8. Constraint plane for the ionization history with $n = -1.0$ (see Fig. 7).

Let us first consider the case of no compact objects. In Fig. 7, we combine the above constraints for four models which bracket the interesting values of $\Omega_0$ and $h$. These models all have the favored spectral index of $n = -0.5$ (Cen, Ostriker, & Peebles 1993). We have also plotted the values of $\sigma_8$ as a function of ionization history. Note that the value of $\sigma_8$ for any spectral index can be read from this figure by using Fig. 7 and equation (1). For example for $\Omega_0 = 0.1$ and $h = 0.5$, $F(z_i = 500, x_i = 0.1) = \sigma_8(n = -0.5, z_i = 500, x_i = 0.1)/\sigma_8^{(NR)}(n = -0.5) = 2.6$. Therefore $\sigma_8(n = -1.0, z_i = 500, x_i = 0.1) \approx 0.45$; whereas the actual value for this model is $\sigma_8 = 0.46$. Notice that if we require $\sigma_8 \approx 1$, the FIRAS Compton-$y$ constraint excludes nearly all models in this range of $\Omega_0$ and $h$. We have also plotted in dashed lines the corresponding constraints if the upper limits on the three relevant experiments, OVRO, ATCA, and FIRAS are brought down by a factor of 2. In the case of the FIRAS Compton-$y$ constraint, this dashed curve is also the constraint for the more realistic value of $T_e = 10000K$ in equation (4). Viable models therefore must have a steeper spectrum than $n \approx -0.5$. Interestingly enough, numerical simulations show that $n \approx -1$ fully ionized models are consistent with large scale structure formation (Suginohara & Suto 1992). In Fig. 8, we plot the same constraints for $n = -1.0$, and we see that reasonable values of $\sigma_8$ fall within the allowed region.

| $(\Omega_0, h)$ | $n$ | $z_i$ | $x_i$ | $\sigma_8$ | TENE | SP91 | PYTH | ARGO | MSAM2 | MAX | MSAM3 |
|---|---|---|---|---|---|---|---|---|---|---|---|
| (0.1,0.5) | −0.5 | 200 | 1 | 1.4 | *1.5* | **2.3** | *2.9* | *2.4* | **3.5** | *3.1* | *1.5* |
| (0.1,0.5) | −1.0 | 50 | 1 | 0.8 | *1.3* | **1.8** | *2.3* | *1.8* | **2.8** | **2.5** | *1.5* |
| (0.1,1.0) | −0.5 | 375 | 0.4 | 1.8 | *1.5* | **2.8** | *3.9* | *3.2* | **5.2** | *4.7* | **2.8** |
| (0.1,1.0) | −1.0 | 50 | 1 | 1.0 | *1.5* | **2.2** | *2.7* | *2.2* | **3.3** | *3.0* | *1.6* |
| (0.2,0.5) | −0.5 | 300 | 0.7 | 1.5 | *1.4* | **2.5** | *3.2* | *2.6* | **4.0** | *3.5* | *1.8* |
| (0.2,0.5) | −1.0 | 50 | 1 | 1.2 | *1.3* | **1.9** | *2.4* | *1.9* | **2.9** | *2.6* | *1.4* |
| (0.2,1.0) | −0.5 | 800 | 0.09 | 2.7 | *1.5* | **2.7** | *3.9* | *3.3* | **5.9** | *5.4* | **3.8** |
| (0.2,1.0) | −1.0 | 250 | 0.5 | 1.2 | *1.4* | **2.4** | *3.1* | *2.5* | **3.9** | *3.5* | *1.9* |
| (0.1,0.5) | −0.5 | 1000 | 0.2 | 1.1 | *1.5* | **2.6** | *3.5* | *2.9* | **4.7** | *4.3* | **2.7** |
| (0.1,0.5) | −0.5 | 500 | 0.4 | 1.0 | *1.5* | **2.5** | *3.2* | *2.7* | **4.2** | *3.8* | *2.1* |
| (0.1,0.5) | −0.5 | 300 | 1 | 1.0 | *1.5* | **2.4** | *3.0* | *2.4* | **3.5** | *3.1* | *1.5* |
| (0.1,1.0) | −1.0 | 1000 | 0.0007 | 0.9 | *1.5* | **3.0** | *4.3* | *3.6* | **6.1** | *5.5* | **3.7** |
| (0.1,1.0) | −1.0 | 300 | 0.04 | 1.0 | *1.4* | **2.4** | *3.1* | *2.5* | **4.2** | *3.8* | **2.5** |
| (0.1,1.0) | −1.0 | 100 | 1 | 0.9 | *1.5* | **2.4** | *3.0* | *2.4* | **3.5** | *3.1* | *1.6* |
| (0.2,0.5) | −1.0 | 1000 | 0.01 | 1.0 | *1.6* | **3.3** | *4.5* | *3.7* | **6.2** | *5.6* | **3.5** |
| (0.2,0.5) | −1.0 | 400 | 0.03 | 1.0 | *1.6* | **2.7** | *3.5* | *2.8* | **4.4** | *4.0* | **2.4** |
| (0.2,0.5) | −1.0 | 100 | 1 | 1.1 | *1.3* | **1.9** | *2.4* | *1.9* | **2.8** | **2.5** | *1.3* |
| (0.2,1.0) | −1.0 | 1000 | 0.05 | 1.0 | *1.5* | **2.3** | *3.1* | *2.5* | **4.3** | *3.9* | **2.7** |
| (0.2,1.0) | −1.0 | 600 | 0.1 | 1.1 | *1.4* | **2.2** | *3.0* | *2.5* | **4.2** | *3.8* | **2.5** |
| (0.2,1.0) | −1.0 | 450 | 0.2 | 1.1 | *1.4* | **2.2** | *3.0* | *2.6* | **4.2** | *3.8* | **2.3** |
| CDM | − | − | − | 1.0 | **0.9** | *1.3* | *1.8* | *1.6* | *2.7* | **2.5** | *1.6* |
| expt | | | | | 1.5±0.3 | 0.9±0.3 | 2.0±0.6 | 1.5±0.2 | 1.7±0.6 | 4.4±0.9 | 1.5±0.4 |
| $\ell$ | | | | | ∼20 | ∼70 | ∼70 | ∼100 | ∼140 | ∼160 | ∼250 |

TAB. 2. Comparison of predicted anisotropies for partially ionized models, $(\Delta T/T)_{rms}$ in units of $10^{-5}$, with observational detections. Bold face numbers show predictions which are in conflict with observation at the 95% CL. The first 8 entries represent models with "minimal" fluctuations at $\ell \sim 100$ of Fig 9. The absolute minimal fluctuations can be slightly smaller ($\sim 10\%$) due to growth between the drag epoch and last scattering (see text). All models are therefore excluded by the SP91 observation at the 95% limit. Models with high $h$ and $n$ are also inconsistent with most other experiments. Latter entries are chosen to be consistent with $\sigma_8 \approx 1$. A standard CDM with $h = 0.5$ and $\Omega_B = 0.06$ is also shown for comparison. For the experimental values, TENE is the quoted rms value; ARGO, MSAM2, MSAM3 are scaled gaussian autocorrelation function results; SP91, PYTH, MAX are effective rms values for an analysis that properly takes into account correlations between data points for a Harrison-Zel'dovich spectrum (White, Scott, Silk 1994).

Now let us consider the radiation fluctuations on degree scales. We have shown in section §2 that the way to minimize fluctuations on the $\ell \lesssim 100$ scale is to maximize $x_i$, minimize $z_i$ and have no compact objects. In Fig. 9, we plot the minimal fluctuations for the eight models of Figs 6 and 7. These models are *all* in conflict with the SP91 13pt scan of Schuster *et al.* (1993) at greater than the 95% CL (see Table 2) and have significantly larger intermediate scale anisotropies than the standard CDM model. Since these models have the minimal fluctuations at the SP91 scale of $\ell \sim 70$, we conclude that if the results of this experiment are confirmed, *all* models regardless of ionization or compact object formation are excluded.

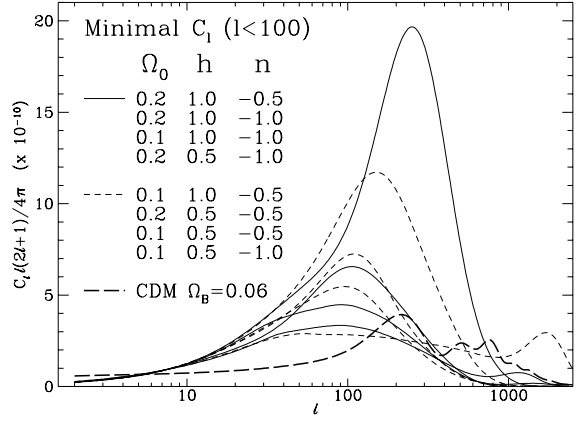

FIG. 9. Minimal temperature anisotropies at $\ell \lesssim 100$. All models are inconsistent with the upper limit from SP91 (see Tab. 2, Schuster *et al.* 1993). Therefore all thermal histories are also inconsistent, including those with compact objects. Fluctuations are also significantly larger than in the standard CDM model (thick dashed line, $h = 0.5$, $\Omega_B = 0.06$) even in these minimal models.

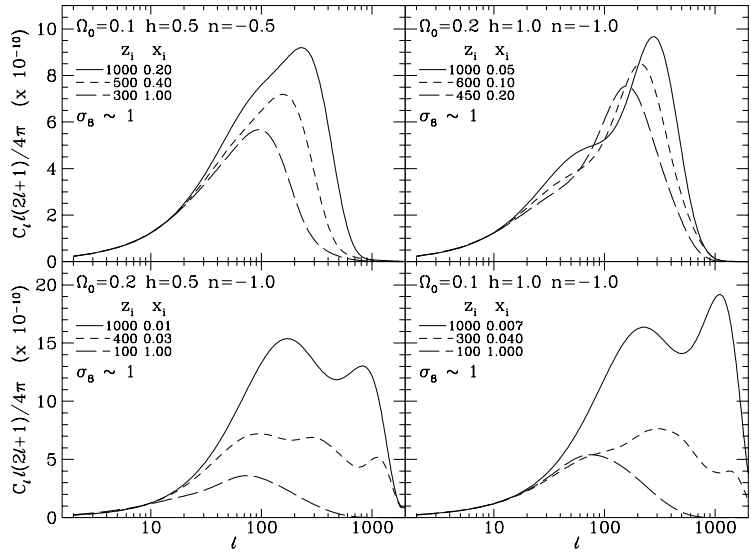

FIG. 10. Temperature anisotropy for partially ionized models with $\sigma_8 \approx 1$. Models with early last scattering $x_i \lesssim 0.01$ overproduce fluctuations since the last scattering surface is nearly at the standard recombination epoch. Corresponding predictions for various experiments are given in Tab. 2.

Unfortunately, agreement between degree scale experiments is marginal at best. In particular, PYTH which is at a very similar angular scale as SP91, detects a somewhat larger fluctuation. Furthermore, MAX GUM (Gundersen et al. 1993) has detected significantly larger fluctuations on a slightly smaller scale. In Table 2, we also compare the predictions of these models with several other experiments. Only $h = 1.0$ and $n = -0.5$ models are excluded at the 95% CL by the PYTH experiment and none are excluded by MAX GUM. In fact, the MAX GUM and TENE results favor isocurvature baryon models over standard CDM. However, only $h = 0.5$ and $n = -1.0$ models satisfy ARGO and MSAM2. Given the current state of confusion, it is perhaps too early to declare isocurvature baryon models excluded on the basis of the SP91 13pt scan (or any individual experiment).

We therefore also show the predictions for realistic models which fit the requirement that $\sigma_8 \approx 1$ (see Fig. 10 and Tab. 2). Models with low ionization fraction $x_i \lesssim 0.01$ tend to overproduce fluctuations on sub-degree scales. Although the optical depth is high enough to erase primary fluctuations, in these models the last scattering surface is so close to $z = 1000$ that equally large fluctuations are generated on the new last scattering surface. For models with a more recent last scattering surface, we obtain values closer to the minimal ones described above.

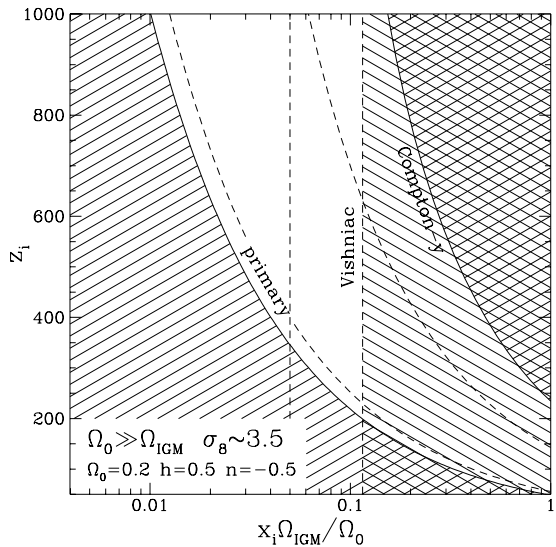

FIG. 11. Example of the modification of the constraint plane (Figs. 7 & 8) to the compact object dominated case. The primary and Compton-$y$ constraints come from optical depth and thus $x_i \to x_i \Omega_{IGM}/\Omega_0$. The Vishniac effect depends additionally on the amplitude of the fluctuations, i.e. $\sigma_8$. Since $\sigma_8$ is fixed to be near the standard recombination value in these models, the Vishniac constraint is the same for a fixed $x_i \Omega_{IGM}/\Omega_0$. The other panels of Figs. 7 & 8 can be generated in a similar way.

In models which are dominated by compact objects, $\sigma_8$ is nearly independent of the ionization history since the baryons fall into the compact object wells by the present. The values for $\sigma_8$ are the same as the standard recombination ones of Fig. 2. Moreover, the constraint diagrams (Figs. 7 and 8) are also easily generalized to the compact object dominated case. Since the primary fluctuation suppression and the Compton-$y$ parameter are purely functions of optical depth, one only needs to replace $x_i$ with $x_i \Omega_{IGM}/\Omega_0$. The Vishniac effect behaves similarly except that it depends sensitively on the magnitude of the fluctuations at last scattering. It is therefore determined from the value of $\sigma_8$, which is fixed in these models, and $x_i$ which determines the epoch of last scattering. An example of a constraint plane is displayed in Fig 11. The other planes may be generated in a similar fashion.

By requiring consistency with $\sigma_8 = 1$, we can fix the spectral index $n$ in compact object dominated universes. We will now take the more general scenario where after compact object formation at $z_c \approx 800$, the universe can be partially ionized, e.g., to $x_c = 0.1$ (Gnedin & Ostriker 1992, Cen, Ostriker, & Peebles 1993).

At the galaxy formation epoch $z_i$, the universe becomes fully ionized $x_i = 1.0$. Again, the constraints can be easily generalized to this model by noting that the primary and Compton-$y$ limits are functions of optical depth. We take $z_i = 5, 20, 50, 800$ as representative cases. The last model corresponds to a universe which was fully ionized at the epoch of compact object formation. In Fig. 12, we plot the anisotropy spectrum; the corresponding predictions for various experiments can be found in Table 3. The models with low secondary ionization redshift $z_i$ overproduce sub-degree scale fluctuations because, either the optical depth is too low to satisfy the constraint on primary anisotropies, or the last scattering surface is so early as to be nearly identical with standard recombination. If $x_c \lesssim 0.1$, only models with $z_i \gtrsim 100$ seem to be viable, in conflict with numerical work which shows $z_i \approx 20 - 30$ in these models (Cen, Ostriker, & Peebles 1993). Notice that these compact object dominated models are nearly always inconsistent with PYTH, ARGO, and MSAM as well as SP91. Should these detections be confirmed, only models with lower small scale fluctuations, *i.e.*, smaller $n$, will survive. Since in order to satisfy $\sigma_8 \approx 1$, $n$ must decrease with $\Omega_0$, this implies that higher $\Omega_0$ values would be desirable. It however remains to be seen whether these low $n$ models are consistent with large scale structure.

| $(\Omega_0, h)$ | $n$ | $z_i$ | $\tau$ | $\sigma_8$ | TENE | SP91 | PYTH | ARGO | MSAM2 | MAX | MSAM3 |
|---|---|---|---|---|---|---|---|---|---|---|---|
| $(0.1, 0.5)$ | $-0.85$ | 5 | 1.4 | 1.0 | *1.5* | **3.3** | **4.8** | **4.1** | **7.2** | **6.6** | **4.7** |
| $(0.1, 0.5)$ | $-0.85$ | 20 | 1.5 | 1.1 | *1.5* | **3.2** | **4.7** | **3.9** | **7.0** | **6.3** | **4.5** |
| $(0.1, 0.5)$ | $-0.85$ | 50 | 1.6 | 1.1 | *1.5* | **2.9** | **4.2** | **3.6** | **6.3** | *5.7* | **4.0** |
| $(0.1, 0.5)$ | $-0.85$ | 800 | 14 | 1.0 | *1.5* | **2.7** | **3.5** | **2.9** | **4.5** | *4.0* | *2.3* |
| $(0.1, 1.0)$ | $-1$ | 5 | 2.9 | 1.0 | *1.5* | **3.0** | **4.2** | **3.5** | **5.8** | **5.2** | **3.4** |
| $(0.1, 1.0)$ | $-1$ | 20 | 3.0 | 1.0 | *1.5* | **2.8** | **3.9** | **3.3** | **5.5** | **4.9** | **3.2** |
| $(0.1, 1.0)$ | $-1$ | 50 | 3.2 | 1.0 | *1.4* | **2.5** | **3.4** | **2.8** | **4.6** | **4.2** | **2.7** |
| $(0.1, 1.0)$ | $-1$ | 800 | 29 | 0.9 | *1.4* | **2.4** | *3.1* | **2.5** | **3.9** | **3.4** | *1.9* |
| $(0.2, 0.5)$ | $-1$ | 5 | 2.1 | 1.0 | *1.6* | **3.3** | **4.6** | **3.9** | **6.4** | **5.8** | **3.7** |
| $(0.2, 0.5)$ | $-1$ | 20 | 2.1 | 1.1 | *1.6* | **3.2** | **4.5** | **3.7** | **6.2** | **5.5** | **3.5** |
| $(0.2, 0.5)$ | $-1$ | 50 | 2.3 | 1.1 | *1.5* | **2.9** | **3.9** | **3.2** | **5.3** | **4.8** | **3.0** |
| $(0.2, 0.5)$ | $-1$ | 800 | 21 | 1.0 | *1.5* | **2.3** | *2.9* | **2.4** | **3.5** | **3.1** | *1.6* |
| $(0.2, 1.0)$ | $-1.15$ | 5 | 4.1 | 1.0 | *1.6* | **2.8** | **3.7** | **3.0** | **4.7** | **4.2** | **2.5** |
| $(0.2, 1.0)$ | $-1.15$ | 20 | 4.2 | 1.0 | *1.5* | **2.6** | **3.5** | **2.8** | **4.4** | **3.9** | **2.3** |
| $(0.2, 1.0)$ | $-1.15$ | 50 | 4.7 | 1.1 | *1.4* | **2.3** | *2.9* | **2.4** | **3.8** | **3.4** | **2.1** |
| $(0.2, 1.0)$ | $-1.15$ | 800 | 41 | 1.0 | *1.4* | **2.1** | *2.7* | **2.2** | **3.4** | **3.0** | *1.6* |
| expt | | | | | 1.5±0.3 | 0.9±0.3 | 2.0±0.6 | 1.5±0.2 | 1.7±0.6 | 4.4±0.9 | 1.5±0.4 |
| $\ell$ | | | | | ~20 | ~70 | ~70 | ~100 | ~140 | ~160 | ~250 |

TAB. 3. Comparison of predicted anisotropies for compact object dominated models, $(\Delta T/T)_{rms}$ in units of $10^{-5}$, which have $n$ chosen to be consistent with $\sigma_8 \approx 1$ (see also Fig. 12). Bold faced values show predictions which are in conflict with observations at the 95% CL. These models have compact object formation at $z_c = 800$ followed by partial ionization $x_c = 0.1$ with $\Omega_{IGM}/\Omega_0 = 0.9$. All models are inconsistent with SP91 at greater than 95% CL. Models with low ionization redshift $z_i$ overproduce fluctuations due to thinness of the last scattering surface and lack of optical depth. Most models are inconsistent with a majority of degree scale experiments. Higher $\Omega_0$ models are more consistent with observations.

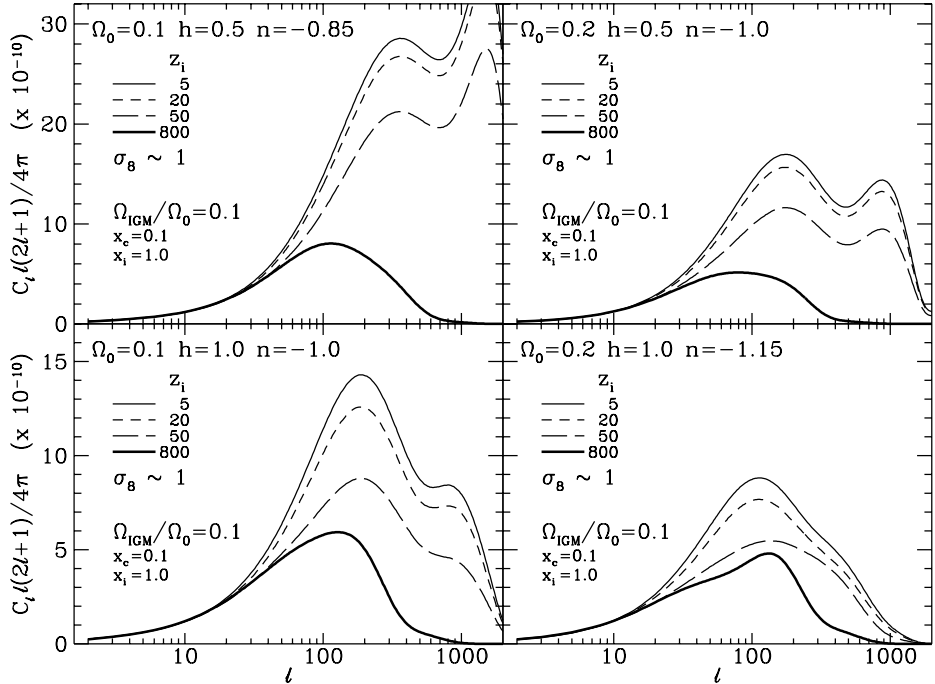

FIG. 12. Temperature anisotropy for compact object dominated models $\Omega_C/\Omega_0 = 0.9$ with $\sigma_8 \approx 1$. The ionization level after compact object formation at $z_c = 800$ is $x_c = 0.1$ until $z_i$ at which point the ionization rises to $x_i = 1.0$. Models with low ionization epoch $z_i$ tend to overproduce fluctuations since they either do not satisfy the optical depth constraint from primary fluctuations $\tau < \tau_{min}$ or have last scattering so early as to regenerate fluctuations as large as standard recombination ones. Predictions for various experiments are given in Tab. 3.

## 4. DISCUSSION

The CMB potentially provides the most serious constraint on the isocurvature baryon model. As we have seen, the COBE DMR detection and galaxy clustering on the $8h^{-1}$ Mpc scale fixes the primordial spectral index for any given ionization history. Furthermore, the ionization history is no longer a free function. It is strongly constrained by spectral distortions, arcminute scale primary anisotropies, and the Vishniac effect. Given these constraints, degree scale anisotropies are the Achilles heel of this model. These anisotropies are significantly larger than their CDM counterparts and are already in conflict with some degree scale experiments. However, indications of a steep slope by COBE and the high detections of TENE and MAX argue in favor of this model. Even if the latter indications are confirmed, it remains to be seen if a fully consistent model can be built. Observational evidence of a smooth, featureless power spectrum (Peacock & Dodds 1994) may conflict with this model. This is especially true for compact object dominated universes where the peak of the matter power spectrum and its small scale oscillatory structure are prominent. Such models can be made more consistent with observations by taking compact object formation to be relatively late $z_c \approx 100$ and assuming *yet another* population of early forming objects which reionize the universe between standard recombination and compact object formation. The matter power spectrum in these models will be similar to our partially ionized ones. Of course, all this fine tuning of the ionization history in order to satisfy microwave background and large scale structure constraints should also arise naturally from physical processes in the model. Nevertheless, relatively large intermediate scale anisotropies is a *robust* prediction of the model. With the expected rapid improvement in these measurements, we shall soon be able to make a definitive statement about this model regardless of uncertainties in the ionization history and other parameters.

## ACKNOWLEDGEMENTS

We would like to acknowledge Douglas Scott and Martin White for detailed discussions of experimental results and for providing the corresponding window functions. We also thank Emory Bunn and Joseph Silk for many useful discussions. N.S. acknowledges financial support from a JSPS Postdoctoral Fellowship for Research Abroad. W.H. was partially supported by the NSF.